\documentstyle[sprocl]{article}

\def\be{\begin{equation}}
\def\ee{\end{equation}}
\def\bea{\begin{eqnarray}}
\def\eea{\end{eqnarray}}

\begin{document}

\title{	SOFT TERMS FROM DILATON/MODULI SECTORS\footnote{talk given at
the International Workshop on ``Elementary Particle Physics Present and
Future'', Valencia (Spain), June 5--9, 1995.}
}

\author{ C. Mu\~noz\footnote{Research supported in part by:
the CICYT, under
contract AEN93-0673; the European Union,
under contracts CHRX-CT93-0132 and
SC1-CT92-0792.}
}

\address{Departamento de F\'{\i}sica Te\'orica C-XI, Universidad Aut\'onoma de
Madrid,\\ Cantoblanco, 28049 Madrid, Spain\\{\sf cmunoz@ccuam3.sdi.uam.es}
}




\maketitle\abstracts{
We study the structure of the soft SUSY-breaking terms obtained
from large classes of 4-D Strings
under the assumption of dilaton/moduli dominance in the process of
SUSY-breaking.
In particular, we first
analyze in detail the
dilaton-dominated limit because of its finiteness properties and
phenomenological predictivity, and second, we consider the new
features appearing when several moduli fields contribute to SUSY breaking.
Although some qualitative
features indeed change in
the multimoduli case with respect to the dilaton dominance one, the
most natural mass relations at low-energy
$m_l < m_q \simeq M_g$ are still similar.
We also study the presence of tachyons pointing out that
their possible existence may be, in some cases,
an interesting advantage in order to break extra gauge symmetries.
Finally, we find that the mechanism
for generating a ``$\mu $-term'' by the K\"ahler potential,
as  naturally implemented in orbifolds,
leads to the prediction $|tg\beta |=1$ at the String scale,
independently of the Goldstino direction.
In this connection, it is worth noticing that in the dilaton-dominated case
we obtain the remarkable result that the whole SUSY spectrum is completely
determined with no free parameters.
}

\rightline{FTUAM 95/30}
\rightline{September 1995}

\section{Introduction}

Recently there has been some activity in trying to obtain information
about the structure
of soft Supersymmetry (SUSY)-breaking
terms in effective $N=1$ theories coming from
four-dimensional Strings. The basic idea is to identify some $N=1$ chiral
fields
whose
auxiliary components could break SUSY by acquiring a vacuum expectation
value (vev).
No special assumption is made about the possible origin of SUSY-breaking.
Natural
candidates in four-dimensional Strings are 1) the complex dilaton field
$S={{1}\over {g^2}}
+ia$ which is present in any four-dimensional String and 2) the moduli fields
$T^i, U^i$ which parametrize the size and shape of the compactified variety
in models obtained by compactification of a ten-dimensional heterotic String.

The important point in this assumption of locating the seed of
SUSY-breaking
in the dilaton/moduli sectors, is that it leads to some interesting
relationships among different soft terms which could perhaps be experimentally
tested.

In ref.\cite{BIM} a systematic discussion of
the structure of soft terms which may be obtained under the assumption of
dilaton/moduli dominated SUSY breaking in some classes of four-dimensional
Strings was presented,
with particular emphasis on the case of Abelian $(0,2)$ orbifold models.
It was mostly considered a situation in which only the dilaton $S$ and
an ``overall modulus $T$'' field contribute to SUSY-breaking.
In fact, actual four-dimensional Strings like orbifolds contain several $T_i$
and $U_i$ moduli.
Thus it is natural to ask what changes if one
relaxes the overall modulus hypothesis and works with the multimoduli
case \cite{BIMS}.
This is one of the purposes of the present talk.
The second one is to analyze in more detail the dilaton-dominated limit, where
only the dilaton field contributes to SUSY breaking. This is a very interesting
possibility not only due to phenomenological reasons, as universality of
the soft terms, but also to theoretical arguments.
In this connection it is worth noticing that
the  boundary conditions $-A=M_{1/2}={\sqrt{3}}m$
of dilaton dominance coincide with some boundary
conditions considered by Jones, Mezincescu and Yau
in 1984 \cite{JMY}. They found that those same boundary conditions
mantain the (two-loop) finiteness properties of
certain $N=1$ SUSY theories.
This could perhaps be an indication that at least
some of the possible soft terms obtained in
the present scheme could have a more general
relevance.

In section 2 we present an analysis of the effects of several moduli
on the results obtained for soft terms.
In the multimoduli case several parameters are needed to specify the
Goldstino direction in the dilaton/moduli space, in contrast with
the overall modulus case where the relevant information is contained
in just one angular parameter $\theta$. The presence of more free
parameters leads to some loss of predictivity for the soft terms.
This predictivity is recovered and increased in the case of dilaton
dominance, where the soft terms eq.(\ref{dilaton})
are independent of the 4-D String
considered and fulfil the low-energy mass relations
given by eq.(\ref{dilaton2}).
Also we show that, even in the multimoduli case, in some schemes
there are certain
sum-rules among soft terms eq.(\ref{rulox}) which hold independently
of the Goldstino direction.
The presence of these sum rules  cause that,
{\it on average} the
{\it qualitative} results in the dilaton-dominated case
still apply. Specifically, if one insists e.g. in
obtaining scalar masses heavier than gauginos
(something not possible
in the dilaton-dominated scenario),
this is possible in the multimoduli case, but
the sum-rules often force some of the scalars
to get negative squared mass.
If we want to avoid this, we have to
stick to gaugino masses bigger than
(or of order) the scalar masses.
This would lead us back to the qualitative results
obtained in dilaton dominance.
In the case of
standard model 4-D Strings this
tachyonic behaviour may be particularly problematic,
since charge and/or colour could be broken.
In the case of GUTs constructed from Strings,
it may just be the signal
of  GUT symmetry breaking.
However, even in this case one expects the same order of
magnitude results for observable scalar and gaugino masses and hence the most
natural mass relations
{\it at low-energy} are still similar to the dilaton dominance ones.

Another topic of interest is the $B$-parameter, the soft mass term
which is associated to a SUSY mass term $\mu H_1H_2$ for the
pair of Higgsses $H_{1,2}$ in the Minimal Supersymmetric Standard Model (MSSM).
Compared to the other soft
terms, the result for the $B$-parameter is more model-dependent.
Indeed, it depends not only on the dilaton/moduli
dominance assumption but also on the particular mechanism which could
generate the associated ``$\mu$-term'' \cite{review}.
An interesting possibility to
generate such a term is the one suggested in ref.\cite{GM}
in which it was
pointed out that in the presence of certain bilinear terms in the
K\"ahler potential an effective $\mu$-term of order the gravitino
mass, $m_{3/2}$, is naturally
generated. Interestingly enough, such bilinear terms in the
K\"ahler potential do appear in String models and particularly in
Abelian orbifolds. In section 3 we compute the $\mu $ and
$B$ parameters\footnote{The results for $B$ corresponding to other sources
for the $\mu$-term can also be found, for the multimoduli case
under consideration, in ref.\cite{BIMS}.
In particular, the possibility of generating a small
$\mu$-term from the superpotential \cite{CM} was studied.}
as well as the soft scalar masses of the
charged fields which could play the role of Higgs particles in
such Abelian orbifold schemes. We find the interesting result that,
{\it independently of the Goldstino direction} in the
dilaton/moduli space, one gets the prediction $|tg\beta |=1$
at the String scale.
On the other hand, if we consider the interesting Goldstino direction
where only the dilaton breaks SUSY, the whole soft terms and the
$\mu$-parameter depend only on the gravitino mass. Imposing the
phenomenological requirement of correct electroweak breaking we arrive
to the remarkable result that the whole SUSY spectrum is completely
determined with no free parameters.

\section{Soft terms}
\subsection{General structure of soft terms: the multimoduli case}
\label{subsec:general}

We are going to consider $N=1$ SUSY 4-D Strings with
$m$ moduli $T_i$, $i=1,..,m$. Such notation refers to both $T$-type
and $U$-type (K\"ahler class and complex structure in the Calabi-Yau
language) fields.
In addition there will be charged matter fields $C_{\alpha }$ and the
complex dilaton field $S$.
In general we will be considering $(0,2)$ compactifications and thus the
charged
fields do not need to correspond to $27$s of $E_6$.

Before further specifying the class of theories that we are going to consider
a comment about the total number of moduli is in order.
We are used to think of large numbers of $T$ and $U$-like moduli
due to the fact that in $(2,2)$ ($E_6$) compactifications there is a
 one to one correspondence between moduli and charged
fields. However, in the case of $(0,2)$ models
with arbitrary gauge group (which is the case of
phenomenological interest) the number of moduli is drastically reduced.
 For example,
in the standard $(2,2)$ $Z_3$ orbifold there are 36 moduli $T_i$,
9 associated to the untwisted sector and 27 to the fixed points of the
orbifold.
In the thousands of $(0,2)$ $Z_3$ orbifolds one can construct by adding
different
gauge backgrounds or doing different gauge embeddings, only the
9 untwisted moduli remain in the spectrum.
The same applies to models with $U$-fields. This is also the case
for compactifications using $(2,2)$ minimal superconformal models. Here all
singlets associated to twisted sectors are projected out when proceeding
to $(0,2)$.
So, as these examples
show, in the case of $(0,2)$ compactifications
 the number of moduli is drastically reduced
to a few fields.
In the case of generic Abelian orbifolds one is in fact left with
only three T-type moduli $T_i$ ($i=1,2,3$), the only exceptions being
$Z_3$, $Z_4$ and $Z'_6$, where such number is 9, 5 and 5 respectively.
The number of $U$-type fields in these $(0,2)$ orbifolds oscillates
between $0$ and $3$, depending on the specific example.
Specifically, $(0,2)$ $Z_2\times Z_2$ orbifolds have 3 $U$ fields,
the orbifolds of type $Z_4,Z_6$,$Z_8,Z_2\times Z_4$,$Z_2\times Z_6$ and
$Z_{12}'$ have just one $U$ field and the rest have no untwisted $U$-fields.
Thus, apart from the three exceptions mentioned above,
this class of models has at most 6 moduli, three of $T$-type (always
present) and at most three of $U$-type. In the case of models obtained
from Calabi-Yau type of compactifications
a similar effect is expected and  only one $T$-field associated to the
overall modulus is guaranteed to exist in $(0,2)$ models.

We will consider effective $N=1$ supergravity (SUGRA)
K\"ahler potentials of the
type:
\begin{eqnarray}
& K(S,S^*,T_i,T_i^*,C_{\alpha},C_{\alpha}^*)\ = \
-\log(S+S^*)\ +\ {\hat K}(T_i,T_i^*)\
\nonumber\\
&+\
{\tilde K}_{{\overline{\alpha }}{ \beta }}(T_i,T_i^*){C^*}^{\overline {\alpha}}
C^{\beta }\
+ (Z_{{\alpha }{ \beta }}(T_i,T_i^*){C}^{\alpha}
C^{\beta }\ +\ h.c. \ ) \ . &
\label{kahl}
\end{eqnarray}
The first piece is the usual term corresponding to the complex
dilaton $S$ which is present for any compactification whereas the
second is the K\"ahler potential of the moduli fields, where we recall
that we are denoting
the $T$- and $U$-type moduli collectively by $T_i$.
The greek indices label the matter fields and their
kinetic term functions are  given by
${\tilde K_{{\overline{\alpha }}{ \beta }}}$ and $Z_{{\alpha }{\beta }}$
to lowest order in the matter fields. The last piece is often forbidden
by gauge invariance in specific models although it may be relevant
in some cases as discussed in section 3.
In this section we are going to consider the case of diagonal metric
both for the moduli and the matter
fields\footnote{An extensive
analysis of the off-diagonal case, including the calculation
of the soft terms and their effects on flavour changing neutral currents
(FCNC),
can be found in ref.\cite{BIMS}.}.
Then ${\hat K}(T_i,T_i^*)$ will be a sum
of contributions (one for each $T_i$), whereas
${\tilde K_{{\overline{\alpha }}{ \beta }}}$ will be taken of the
diagonal form ${\tilde K_{{\overline{\alpha }}{ \beta }}}
\equiv \delta _{{\overline{\alpha }}{ \beta }} {\tilde K_{\alpha }}$.
The complete $N=1$ SUGRA Lagrangian is determined by
the K\"ahler potential $K({\phi }_M ,\phi^*_M)$, the superpotential
$W({\phi }_M)$ and the gauge kinetic functions
$f_a({\phi }_M)$, where $\phi_M$ generically denotes the chiral fields
$S,T_i,C_{\alpha }$. As is well known, $K$ and $W$ appear in the
Lagrangian only in the combination $G=K+\log|W|^2$. In particular,
the (F-part of the) scalar potential is given by
\begin{equation}
V(\phi _M, \phi ^*_M)\ =\
e^{G} \left( G_M{K}^{M{\bar N}} G_{\bar N}\ -\ 3\right) \ ,
\label{pot}
\end{equation}
where $G_M \equiv \partial_M G \equiv \partial G/ \partial \phi_M$
and $K^{M{\bar N}}$ is the inverse of the K\"ahler metric
$K_{{\bar N }M}\equiv{\partial}_{\bar N}{\partial }_M K$.

The crucial assumption  now is  to locate the origin of SUSY-breaking in the
dilaton/moduli sector.
Let us take the following parametrization
for the vev's of the dilaton and moduli auxiliary
fields $F^S=e^{G/2} G_{ {\bar{S}} S}^{-1} G_S$ and
$F^i=e^{G/2} G_{ {\bar{i}} i}^{-1} G_i$:
\begin{equation}
G_{ {\bar{S}} S}^{1/2} F^S\ =\ \sqrt{3}m_{3/2}\sin\theta e^{i\gamma _S}\ \ ;\ \
G_{ {\bar{i}} i}^{1/2} F^i\ =\ \sqrt{3}m_{3/2}\cos\theta\ e^{i\gamma _i}
\Theta _i \ \ ,
\label{auxi}
\end{equation}
where $\sum _i \Theta _i^2=1$ and $e^G=m^2_{3/2}$ is the gravitino
mass-squared.
The angle $\theta $ and the $\Theta _i$ just parametrize the
direction of the goldstino in the $S,T_i$ field space.
 We have also allowed for the possibility of
some complex phases $\gamma _S, \gamma _i$ which could be relevant
for the CP structure of the theory. This parametrization has the virtue that
when we plug it in the general form of the SUGRA scalar potential
eq.(\ref{pot}), its vev (the cosmological constant) vanishes by
construction. Notice that such a phenomenological approach allows us
to `reabsorb' (or circumvent) our ignorance about the (nonperturbative)
$S$- and $T_i$- dependent part of the superpotential, which is
responsible for SUSY-breaking.
It is now a straightforward
exercise
to compute the bosonic soft SUSY-breaking terms in this class of theories.
Plugging
eqs.(\ref{auxi}) and (\ref{kahl}) into eq.(\ref{pot})
one finds the following results (we recall that we
are considering here a diagonal metric for the matter fields):
\begin{eqnarray}
 & m_{\alpha }^2 = \  m_{3/2}^2 \ \left[ 1\ -\ 3\cos^2\theta \
({\hat K}_{ {\overline i} i})^{-1/2} {\Theta }_i e^{i\gamma _i}
(\log{\tilde K}_{\alpha })_{ {\overline i} j}
({\hat K}_{ {\overline j} j})^{-1/2} {\Theta }_j e^{-i\gamma _j} \ \right] \ ,
&
\nonumber \\
 & A_{\alpha \beta \gamma } =
 \   -\sqrt{3} m_{3/2}\ \left[ e^{-i{\gamma }_S} \sin\theta \right.
& \nonumber \\
& \left. - \ e^{-i{\gamma }_i} \cos\theta \  \Theta_i
({\hat K}_{ {\overline i} i})^{-1/2}
\ \left(  {\hat K}_i - \sum_{\delta=\alpha,\beta,\gamma}
(\log {\tilde K}_{\delta })_i
+ (\log h_{\alpha \beta \gamma } )_i \ \right)
\  \right] \ . &
\label{soft}
\end{eqnarray}
The above scalar masses and trilinear scalar couplings correspond
to charged fields which have already been canonically normalized.
Here $h_{\alpha \beta \gamma }$ is a renormalizable
Yukawa coupling involving three charged chiral fields and
$A_{\alpha \beta \gamma }$ is its corresponding trilinear soft term.

Physical gaugino masses $M_a$ for the canonically normalized gaugino fields
are given by $M_a=\frac{1}{2}(Re f_a)^{-1}e^{G/2}{f_a}_M
{K}^{M{\bar N}} G_{\bar N}$.
Since the tree-level gauge kinetic function is given for any 4-D String by
$f_a=k_aS$, where $k_a$ is the Kac-Moody level of the gauge factor,
the result for tree-level gaugino masses is independent of the
moduli sector and is simply given by:
\begin{equation}
M\equiv M_a\ =\ m_{3/2}\sqrt{3} \sin\theta e^{-i\gamma _S} \ .
\label{gaugin}
\end{equation}

The soft term formulae above are in general valid for any compactification
as long we are considering diagonal metrics. In addition one is tacitally
assuming that the tree-level K\"ahler potential and $f_a$-functions
constitute a good aproximation.
The K\"ahler potentials
for the moduli are in general complicated functions.
Before going into specific classes of Superstring models, it is worth
studying the interesting limit
$\cos\theta =0$, corresponding to the case where the dilaton sector
is the source of all the SUSY-breaking (see eq.(\ref{auxi})).

\subsection{The $\cos\theta =0$ (dilaton-dominated) limit}
\label{subsec:dilaton}

Since the dilaton couples in an universal manner to all particles,
{\it this limit is quite model independent}. Using
eqs.(\ref{soft},\ref{gaugin}) and neglecting phases
one finds the following simple expressions for the soft terms which are
independent of the 4-D String considered
\begin{eqnarray}
 & m_{\alpha } = \  m_{3/2} \ , &
\nonumber \\
 & A_{\alpha \beta \gamma } =
 \   -\sqrt{3} m_{3/2} \ , &
\nonumber \\
 & M_a =
 \   \sqrt{3} m_{3/2}  \ . &
\label{dilaton}
\end{eqnarray}

This dilaton-dominated scenario \cite{KL,BIM}
is attractive for its simplicity and for
the natural explanation that it offers to the universality of the
soft terms. Actually, universality is a desirable property not
only to reduce the number of independent parameters in the MSSM, but also
for phenomenological reasons, particularly to avoid FCNC.

Because of the simplicity of this scenario, the low-energy predictions
are quite precise. Since scalars are lighter than gauginos
at the String scale, at low-energy ($\sim M_Z$), gluino, slepton and
(first and second generation) squark mass relations turn out to
be \cite{BIM,BLM}
\begin{eqnarray}
M_g:m_Q:m_u:m_d:m_L:M_e  \simeq 1:0.94:0.92:0.92:0.32:0.24 \ .
\label{dilaton2}
\end{eqnarray}
Although squarks and sleptons have the same soft mass, at low-energy the
former are much heavier than the latter because of the gluino contribution to
the renormalization of their masses.

In section 3 we will show that even a stronger result than that of
eq.(\ref{dilaton2}) is obtained in the context of a natural mechanism
for solving the $\mu$ problem, namely the whole SUSY spectrum (gluino,
squarks, sleptons, Higgses, charginos, neutralinos) is completely
determined with no free parameters.

\subsection{Orbifold compactifications}\label{subsec:orbifold}

To illustrate some general features of the multimoduli case
 we will concentrate here on the case of generic $(0,2)$ symmetric
Abelian orbifolds. As we mentioned above, this class of models
contains three $T$-type  moduli and (at most) three $U$-type moduli.
We will denote them collectively by $T_i$, where e.g. $T_i=U_{i-3}$; $i=4,5,6$.
For this  class of models the K\"ahler potential has the form
\begin{equation}
K(\phi,\phi^*)\ =\ -\log(S+S^*)\ -\ \sum _i \log(T_i+T_i^*)\ +\
\sum _{\alpha } |C_{\alpha }|^2 \Pi_i(T_i+T_i^*)^{n_{\alpha }^i} \ .
\label{orbi}
\end{equation}
Here $n_{\alpha }^i$ are fractional numbers usually called ``modular weights"
of the matter fields $C_{\alpha }$. For each given Abelian orbifold,
independently of the gauge group or particle content, the possible
values of the modular weights are very restricted. For a classification of
modular weights for all Abelian orbifolds see ref.\cite{BIMS}.
Using the particular form (\ref{orbi}) of the K\"ahler potential and
eqs.(\ref{soft},\ref{gaugin}) we obtain
the following results\footnote{This analysis was also carried out, for the
particular case of the three diagonal moduli $T_i$,
in ref.\cite{japoneses}
in order to obtain unification of gauge coupling constants.
Some particular multimoduli examples were also considered in
ref.\cite{FKZ}.} for the scalar masses, gaugino masses and soft trilinear
couplings:
\begin{eqnarray}
   &m_{\alpha }^2 =  \  m_{3/2}^2(1\ +\ 3\cos^2\theta\ {\vec {n_{\alpha }}}.
{\vec {\Theta ^2}}) \ , &
\nonumber\\
&  M = \  \sqrt{3}m_{3/2}\sin\theta e^{-i{\gamma }_S} \ , &
\nonumber\\
 & A_{\alpha \beta \gamma } = \   -\sqrt{3} m_{3/2}\ ( \sin\theta e^{-i{\gamma
}_S}
\ +\ \cos\theta \sum _{i=1}^6 e^{-i\gamma _i}    {\Theta }^i {\omega
}^i_{\alpha
\beta \gamma } ) \ , &
\label{masorbi}
\end{eqnarray}
where we have defined :
\begin{equation}
{\omega }^i_{\alpha \beta \gamma }\ =\ (1+n^i_{\alpha }+n^i_{\beta
}+n^i_{\gamma
}-
 {Y}^i_{\alpha \beta \gamma }    )\ \ ;\ \
{Y}^i_{\alpha \beta \gamma } \
= \ {{h^i_{\alpha \beta \gamma }}\over {h_{\alpha \beta \gamma
}}} 2ReT_i \ .
\label{formu}
\end{equation}
Notice that neither the scalar nor the gaugino masses have any explicit
dependence on $S$ or $T_i$, they only depend on the gravitino mass and
the goldstino angles.
This is one of the advantages of a parametrization in terms of such angles.
Although in the case of the $A$-parameter
an explicit $T_i$-dependence may appear in
the term proportional to $Y^i_{\alpha \beta \gamma }$, it disappears in
several interesting cases \cite{BIMS}.

With the above information we can now analyze the different structure of
soft terms available for each Abelian orbifold.

{\it 1) Universality of soft terms}

In the dilaton-dominated case ($\cos\theta =0$) the whole
soft terms are universal.
However, in general, they show a lack of universality due to the
modular weight dependence (see eqs.(\ref{masorbi},\ref{formu})).

{\it 2) Soft masses}

In the
multimoduli case, depending on the goldstino direction, tachyons
may appear. For $\cos^2\theta \geq 1/3 $, one has to
be very careful with the goldstino direction if one is interested
in avoiding tachyons.
Nevertheless, as we will discuss below, having a tachyonic sector is
not necessarily a problem, it may even be an advantage, so one should not
disregard this possibility at this point.

Consider now three particles
$C_{\alpha }$,$C_{\beta }$,$C_{\gamma }$
coupling through a Yukawa $h_{\alpha \beta \gamma }$. They may belong
both
to the untwisted (${\bf U}$) sector or to a twisted
(${\bf T}$) sector, i.e. couplings
of the type ${\bf U}{\bf U}{\bf U}$,
${\bf U}{\bf T}{\bf T}$,
${\bf T}{\bf T}{\bf T}$. Then, using the above formulae, one
finds \cite{BIMS}
that in general for {\it any choice} of goldstino direction
\begin{equation}
m_{\alpha }^2\ +\ m_{\beta }^2\ +\ m_{\gamma }^2\ \leq \ |M|^2\
=3 m_{3/2}^2\sin^2\theta \
\label{rulox}
\end{equation}

Notice that if we insist in having a vanishing gaugino mass, the sum-rule
(\ref{rulox}) forces
the  scalars to be either all massless or at least one of them tachyonic.
Nevertheless we should not forget that tachyons, as we already mentioned
above, are not necessarily a problem, but may just show us an instability.

{\it 3) Gaugino versus scalar masses}

In the multimoduli case
on average the scalars
are lighter than
gauginos but there may be scalars with mass bigger than gauginos.
Eq.(\ref{rulox})\ tells us that this can only be true
at
the cost of
having some of the other three scalars with {\it negative} squared mass.
This may have diverse phenomenological
implications depending what is the particle content
of the model, as we now explain in some detail:

{\it 3-a) Gaugino versus scalar masses in standard model 4-D Strings}

Let us suppose we
insist in
e.g., having tree-level gaugino masses lighter than
the scalar masses.
If we are dealing with a String model with gauge group
$SU(3)_c\times SU(2)_L\times U(1)_Y$$\times G$
this is potentially a disaster.  Some
observable particles, like Higgses, squarks or sleptons would be forced
to acquire large vev's (of order the String scale). For example, the scalars
associated through the Yukawa coupling $H_2Q_Lu_L^c$, which generates the
mass of the $u$-quark, must fulfil the above sum-rule
(\ref{rulox}). If we
allow e.g. the scalars $H_2$, $Q_L$ to be heavier than gauginos, then
$u_L^c$ will become tachyonic breaking charge and color.
However, tachyons may be helpful if the particular Yukawa coupling
does not involve observable particles. They could break extra gauge symmetries
and generate large masses for extra particles. We recall that standard-like
models in Strings usually have too many extra particles and many extra
U(1) interactions. Although the Fayet-Iliopoulos mechanism helps to cure
the problem \cite{suplemento}, the existence of tachyons is a complementary
solution.

We thus see that, for standard model Strings,
if we want  to avoid charge and colour-breaking minima (or vev's of
order the String scale for the Higgses),
we should grosso modo come back to a situation
with gauginos heavier than scalars.
Thus the low-energy phenomenological predictions of
the multimoduli case are similar to
those of the
dilaton-dominated scenario (see subsect.2.2):
due to the
sum-rule
the tree-level observable scalars are always ligther than gauginos
\begin{equation}
m_{\alpha} < M \ .
\label{masas1}
\end{equation}
Now, at low-energy ($\sim M_Z$), gluino, slepton and (first and second
generation) squark mass relations turn out to be
\begin{equation}
m_l < m_q \simeq M_g \ ,
\label{masas2}
\end{equation}
where gluinos are slightly heavier than scalars. This
result is qualitatively similar to the dilaton dominance one, in
spite of the different set of (non-universal) soft scalar masses, because
the low-energy scalar masses are mainly determined by the gaugino loop
contributions. The only exception are the sleptons masses, which do not
feel the important gluino contribution, and therefore can get some
deviation from the result of eq.(\ref{dilaton2}).

Although String loop corrections, in the particular case that at
tree-level $M \rightarrow 0$
and $m_{\alpha}\rightarrow 0$,
can yield scalars heavier than gauginos \cite{BIM}, it was shown in
ref.\cite{BIMS} that this possibility is a sort of fine-tuning.
Non-renormalizable Yukawa couplings can also yield scalars heavier
than gauginos. However
it was shown in ref.\cite{BIMS} that
still at low-energy eq.(\ref{masas2}) is
typically valid, the only
difference being that now squarks will be slightly heavier than gluinos.

{\it 3-b ) Gaugino versus scalar masses in GUT 4-D Strings}

What it turned out to be a potential disaster
in the case of standard model Strings may be an interesting
advantage in the case of String-GUTs.
In this case it could well be that
the negative squared mass may
just induce gauge symmetry breaking by forcing a vev for a particular
scalar (GUT-Higgs field) in the model.
The latter possibility provides us with interesting phenomenological
consequences.
Here the breaking of SUSY would directly induce further gauge symmetry
breaking.
An explicit example of this situation can be found in ref.\cite{BIMS}.

Let us finally remark that, in spite of the different possibilities of soft
masses in the multimoduli case, the most natural (slepton-squark-gluino)
mass relations
{\it at low-energy} will be similar to the ones of
the dilaton-dominated case eq.(\ref{masas2}) as showed in point {\it 3-a}.

\section{The B-parameter and the $\mu $ problem}

It was pointed out in ref.\cite{GM} that terms in a K\"ahler potential
like the one proportional to $Z_{\alpha \beta }$ in eq.(\ref{kahl})
can naturally induce a $\mu $-term for the $C_{\alpha }$ fields
of order $m_{3/2}$ after SUSY-breaking, thus providing a rationale
for the size of $\mu $. Recently it has been suggested that such type of
terms may appear in the K\"ahler potential of some Calabi-Yau type
compactifications \cite{KL}.
It has also been explicitly shown \cite{LLM}
that the untwisted sector of orbifolds
with at least one complex-structure field  $U$  possesses the required
structure $Z(T_i,T_i^*)C_1C_2+h.c.$ in their K\"ahler
potentials. Specifically, the $Z_N$ orbifolds
based on $Z_4,Z_6$,$Z_8,Z_{12}'$ and the $Z_N\times Z_M$ orbifolds based
on $Z_2\times Z_4$ and $Z_2\times Z_6$ do all have a $U$-type field in (say)
the third complex plane. In addition the $Z_2\times Z_2$ orbifold has $U$
fields in the three complex planes.
In all these models the piece of the K\"ahler potential involving
the moduli and the untwisted matter fields $C_{1,2}$ in the third complex
plane has the form
\begin{eqnarray}
K = - \log(T_3+T_3^*)  - \log(U_3+U_3^*)\ +
\frac{(C_1+C_2^*)(C_1^*+C_2)}{(T_3+T_3^*)(U_3+U_3^*)}
\label{kahlexp}
\end{eqnarray}
where one can easily identify the
functions $Z, {\tilde K}_1, {\tilde K}_2$ associated to $C_1$ and $C_2$:
\begin{equation}
Z\ =\ {\tilde K}_1 \ =\ {\tilde K}_2\ =\  {1\over {(T_3+T_3^*)(U_3+U_3^*)}}
\ .
\label{zzz}
\end{equation}
Plugging back these expressions in eqs.(\ref{pot},\ref{auxi})
one can compute $\mu$ and $B$ for this interesting class
of models \cite{BIMS}:
\begin{eqnarray}
& \mu \ =\ m_{3/2}\ \left( 1\ +\ \sqrt{3}\cos\theta
(e^{i \gamma_3} \Theta _3 + e^{i \gamma_6} \Theta _6)\right) \ , &
\label{muu}
\\
& B\mu=2m_{3/2}^2 \left( 1+\sqrt{3} \cos\theta
 ( \cos\gamma_3 \Theta_3 + \cos\gamma_6 \Theta_6)  \  \right.
& \nonumber\\
& \left.
+\ 3\cos^2\theta \cos(\gamma_3-\gamma_6) {\Theta _3}{\Theta _6} \right) \ . &
\label{bmu}
\end{eqnarray}
In addition, we recall from eq.(\ref{masorbi}) that the soft masses are
\begin{equation}
m^2_{C_1}\ =\ m^2_{C_2}\ =\  m_{3/2}^2\ \left( 1\ -\ 3\cos^2\theta
(\Theta_3^2+\Theta _6^2)\right) \ .
\label{mundos}
\end{equation}
In general, the dimension-two scalar potential for $C_{1,2}$
after SUSY-breaking has
the form
\begin{equation}
 V_2(C_1,C_2)\ =\ (m_{C_1}^2+|\mu|^2)|C_1|^2 + (m_{C_2}^2+|\mu| ^2)|C_2|^2
+(B\mu C_1C_2+h.c.)\
\label{flaty}
\end{equation}
In the specific case under consideration,
from
eqs.(\ref{muu},\ref{bmu},\ref{mundos}) we find the remarkable result
that the three coefficients in $V_2(C_1,C_2)$ are equal, i.e.
\begin{equation}
m_{C_1}^2+|\mu|^2 = m_{C_2}^2+|\mu| ^2 = B\mu
\label{result}
\end{equation}
Although the common value of the three coefficients in eq.(\ref{result})
depends on the Goldstino direction via the parameters
$\cos\theta$, $\Theta_3$, $\Theta_6$,\ldots (see expression of $B\mu$
in eq.(\ref{bmu})), we stress that the equality itself
holds {\em independently of the Goldstino direction}.


It is well known that, for a potential of the generic form
(\ref{flaty}) (+D-terms), the minimization conditions yield
\begin{equation}
\sin2\beta  \ =\ { {-2 B\mu} \over {m_{C_1}^2+m_{C_2}^2+2|\mu|^2} } \ .
\label{sbet}
\end{equation}
In particular, this relation embodies the boundedness requirement:
if the absolute value of the right-hand side becomes bigger than one,
this would indicate that the potential becomes unbounded from below.
As we have seen, in the class of models under consideration
the particular expressions of the mass parameters lead to
the equality (\ref{result}), which in turns implies
$\sin 2\beta= -1$. Thus one finds $\tan\beta=<C_2>/<C_1>=-1$
{\it for any value of $\cos\theta $,$\Theta _3 $,$\Theta _6 $} (and of
the other $\Theta_i$'s of course), i.e. for any Goldstino direction.

As an additional comment, it is worth recalling that in previous
analyses of the above mechanism for generating $\mu$ and $B$
in the String context \cite{KL,BIM,BLM} the value of $\mu$ was left
as a free parameter since one did not have an explicit expression for
the function $Z$. However, if the explicit orbifold formulae for
$Z$ are used, one is able to predict both \cite{BIMS} $\mu$ and $B$ reaching
the above conclusion\footnote{We should add
that situations are conceivable
where the above result may be evaded, for example if the physical Higgs
doublets are a mixture of the above fields with some other doublets coming
from other sectors (e.g. twisted) of the theory.}.

Now that we have computed explicitly the whole soft terms and the $\mu$
parameter, it would be interesting to analyze the dilaton-dominated scenario
($\cos\theta=0$) because of its predictivity. In particular, from
eqs.(\ref{dilaton},\ref{muu},\ref{bmu}) we obtain
\begin{eqnarray}
 & m_{\alpha } = \  m_{3/2} \ , &
\nonumber \\
 & A_{\alpha \beta \gamma } =
 \   -\sqrt{3} m_{3/2}  \ , &
\nonumber \\
 & M_a =
 \   \sqrt{3} m_{3/2}   \ , &
\nonumber \\
& B = \  2 m_{3/2} \ , &
\nonumber \\
& \mu = \  m_{3/2} \ . &
\label{dilaton3}
\end{eqnarray}
and therefore the whole SUSY spectrum depends only on one parameter
($m_{3/2}$).
If we would know the particular mechanism which breaks SUSY, then we would
be able of computing the superpotential and hence $m_{3/2}=e^K|W|$. However,
still this parameter can be fixed from the phenomenological requirement
of correct electroweak breaking $2M_W/g_2^2=<H_1>^2+<H_2>^2$. Thus at the
end of the day we are left with no free parameters. The whole SUSY spectrum
is completely determined in this scenario.
This is a remarkable result which deserves further
investigation \cite{nuevo}. Of course, if in the next future the mechanism
which breaks SUSY is known (i.e. $m_{3/2}$ can be explicitly calculated)
and the above scenario is the correct one, the
value of $m_{3/2}$ should coincide
with the one obtained from the phenomenological
constraint.

It is worth noticing here that although the value of $\mu$ is
compactification dependent even in this dilaton-dominated scenario
$\mu=m_{3/2}({\tilde K}_1{\tilde K}_2)^{-1/2}Z$, the result of
eq.(\ref{dilaton3}), $\mu=m_{3/2}$, will be obtained in any compactification
scheme with the following property: ${\tilde K}_1={\tilde K}_2=Z$.
Of course, this is the case of
orbifolds as was shown in eq.(\ref{zzz}).


\section*{References}

\begin{thebibliography}{99}
%
\bibitem{BIM} A. Brignole, L.E. Ib\'{a}\~{n}ez and C. Mu\~noz,
{\it Nucl. Phys.} {\bf B422} (1994) 125 [Erratum: {\bf B436} (1995) 747].
%
\bibitem{BIMS} A. Brignole, L.E. Ib\'{a}\~{n}ez and C. Mu\~noz and C. Scheich,
{\it FTUAM} {\bf 95/26}, {\it hep-ph/} {\bf 9508258}.
%
\bibitem{JMY} D.R.T. Jones, L. Mezincescu and Y.P. Yao,
{\it Phys. Lett.} {\bf B148} (1984) 317.
%
%
%
\bibitem{review} For a recent review, see: C. Mu\~noz, {\it FTUAM} {\bf 95/20},
{\it hep-th/} {\bf 9507108}.
%
\bibitem{GM} G.F. Giudice and A. Masiero, {\it Phys. Lett.} {\bf B206}
(1988) 480.
%
\bibitem{CM} J.A. Casas and C. Mu\~noz, {\it Phys. Lett.} {\bf B306}
(1993) 288.
%
%
%
\bibitem{KL} V.S. Kaplunovsky and J. Louis
{\it Phys. Lett.} {\bf B306} (1993) 269.
%
\bibitem{BLM} R. Barbieri, J. Louis and M. Moretti,
{\it Phys. Lett.} {\bf B312} (1993) 451.
%
\bibitem{japoneses} T. Kobayashi, D. Suematsu, K. Yamada and Y. Yamagishi,
{\it Phys. Lett.} {\bf B348} (1995) 402.
%
%
\bibitem{FKZ} S. Ferrara, C. Kounnas and F. Zwirner,
{\it Nucl. Phys.} {\bf B429} (1994) 589 [Erratum: {\bf B433} (1995) 255].
%
%
%
\bibitem{suplemento} For a review, see e.g.: J.A. Casas and C. Mu\~noz
{\it Nucl. Phys. (Proc. Suppl.)} {\bf B16} (1990) 624, and references
therein.
%
%
%
%
%
\bibitem{LLM} G. Lopes-Cardoso, D. L\"ust and T. Mohaupt,
{\it Nucl. Phys.} {\bf B432} (1994) 68;
I. Antoniadis, E. Gava, K.S. Narain and T.R. Taylor,
{\it Nucl. Phys.} {\bf B432} (1994) 187.
%
\bibitem{nuevo} A. Brignole, L.E. Iba\~nez and C. Mu\~noz, to appear.
%


\end{thebibliography}
\end{document}